\documentstyle[aps,12pt]{revtex}
\newcommand{\mbd}[1]{\mbox{\bf #1}}
\newcommand{\br}{\mbd{r}}
\newcommand{\bx}{\mbd{x}}
\newcommand{\by}{\mbd{y}}
\newcommand{\ba}{\mbd{a}}
\newcommand{\bb}{\mbd{b}}
\newcommand{\bA}{\mbd{A}}
\newcommand{\bi}{\mbd{i}}
\newfont{\mbld}{cmbsy10 scaled 1000}
\begin{document}
\title{TOPOLOGICAL\ ENTANGLEMENT\ OF\ POLYMERS\ AND\ CHERN--SIMONS\
FIELD\ THEORY}
\author{Franco Ferrari \thanks{The work of F. Ferrari
has been supported by the European Union, TMR Programme, under
grant ERB4001GT951315}$^{,a}$  and Ignazio Lazzizzera
$^{b,c}$\\
$^a$ {\it LPTHE\thanks{Laboratoire associ\'e No. 280 au CNRS} ,
Universit\'e Pierre et Marie Curie--PARIS VI and}\\
{\it Universit\'e Denis Diderot--Paris VII,
Boite $126$, Tour 16, $1^{er}$ \'etage,}\\
{\it 4 place Jussieu,
F-75252 Paris CEDEX 05, FRANCE.}\\
$^b$ {\it Dipartimento di Fisica, Universit\'a di Trento, 38050 Povo (TN),
Italy.}\\
$^c$ {\it INFN, Gruppo Collegato di Trento, Italy.}}
\date{February 98}
\maketitle
\vspace{-4.5in} \hfill{Preprint PAR-LPTHE 98--15, UTF 410/98} \vspace{4.4in}
\begin{abstract}
In recent times some interesting field theoretical descriptions
of the statistical mechanics of entangling polymers
have been proposed by various authors. In these approaches,
a single test polymer fluctuating in a background of static
polymers or in a lattice of obstacles
is considered.
The extension to the case in which the configurations of
two or more polymers become non-static is not straightforward unless
their trajectories are severely
constrained.
In this paper we present another approach,
based on Chern--Simons field theory, which is able to
describe the topological entanglements
of two fluctuating  polymers in terms of gauge fields
and second quantized replica fields.

\end{abstract}
\section{Introduction}
In recent times, some interesting field theoretical descriptions of the
statistical mechanics of polymer rings subjected to
topological constraints have been proposed by various
authors
\cite{polftmft}--\cite{kleinert}.
In all these approaches, which are based on the pioneering
works \cite{edwards},\cite{degennes},
the inclusion of higher order link
invariants \cite{kauffmann},\cite{witten},
\cite{jones}
is yet an unsolved problem. On the other side, these invariants are
necessary in
order to specify in an unique way the distinct topological states of the
polymers. The main difficulty in the application
of higher order link invariants is that they
cannot be easily expressed in terms of the variables which
characterize the polymer, i. e. its trajectory
in the three dimensional $(3-D)$ space and its contour length.
For this reason, in all analytical methods used to study entangling polymers
only
the simplest topological invariant is considered,
namely
the Gauss linking number
\cite{polftcsprop}.
Even in this approximation, a rigorous treatment of the $3-D$
entanglement problem is mathematically difficult. Basically,
only the case of a single polymer fluctuating in a background
of static polymers or fixed obstacles has been investigated until now.
The configurations of the background polymers may be ``averaged''
exploiting mean-field type arguments
(see e. g. \cite{polftmft}).

As it has been recently suggested \cite{polftcsprop},\cite{kleinert},
a possible way out from the above difficulties
is the
introduction
of a Chern-Simons (C--S) field theory
\cite{chernsimons}
in the treatment of polymer entanglement.
In doing this, however, one is faced with the problem
of computing expectation values of holonomies around the closed
trajectories formed by the polymers in space.
As it is is well known, these expectation values
are affected
by the presence of ambiguous contributions
coming from the self-linking of the loops.
In pure C--S field theories, a good regularization for these non-topological
terms is
provided by the so-called framing procedure
\cite{witten}.
However, the framing depends on the form of the loop,
which in our case is a dynamical variable, in the sense that
one has to sum over all
the possible
trajectories of the polymers in $3-D$ space. Therefore, the addition of
a framing complicates
the form of the polymer action to the extent that the integration over the
polymer configurations can no longer be performed in any closed form.

To avoid this obstacle, we propose in this paper a model in which the polymers
are coupled to
two abelian C--S fields. The coupling
constants are suitably chosen in order to
cancel all the undesired terms coming
from the self-linking of the loops without the need of a framing.
Of course, this procedure does not replace framing, but it is
sufficient to eliminate the self-linking ambiguities
at least for the C--S amplitudes
that are necessary in our context.
The advantage of introducing auxiliary C--S fields is that
they mediate
the topological interactions between the polymers decoupling their
actions.
As a consequence, each polymer can be separately
treated with the powerful methods developed in refs.
\cite{gaugemod},\cite {tanaka}.
In this way, we are able to formulate
the problem of two fluctuating polymers
subjected to topological constraints in terms of a
C--S gauge field theory with $n$
components in the limit in which $n$ goes to zero.
With respect to refs. \cite{gaugemod},\cite {tanaka}, the only complication
in our approach is that
the external magnetic field arising
due to the presence of the background polymers is replaced by quantum
C--S fields. For instance, we can
compute the analogous of
all polymer configurational
probabilities derived in ref. \cite{tanaka}. This will be done in
Section 3.

The material presented in this paper is divided as follows.
In Section 2 we briefly review the field theoretical approach
developed in refs. \cite{gaugemod},\cite {tanaka} in the case
of a test polymer entangling with another polymer
of fixed conformation.
In Section 3 this approach will be extended to the situation in which both
polymers are dynamical.
Finally, in the Conclusions some possible generalizations and applications
of our treatment will
be discussed.
\section{Statistical--Mechanical Theory of Polymer Entanglement}
Let $P$ be a polymer (see e. g.
ref. \cite{kleinert} for a general introduction
on the physics of polymers) represented as a long chain of $N+1$ segments
$\vec{r}_{i+1} - \vec{r}_i$ for $i=0,\ldots,N$. Each segment
has fundamental step length $a$, which is supposed to be very small with
respect to the total length of the polymer.
Moreover, the junction between adjacent segments is such that they can freely
rotate in all directions. In the limit of large values of $N$, the ensemble
of $M$ polymers $P_1,\ldots, P_M$ of this kind can be regarded as the
ensemble of $M$ particles subjected to a self-avoiding randow walk.
The whole configuration of a polymer $P$ is thus entirely specified by the
trajectory $\br(s)$ of a particle in three dimensional ($3-D$) space,
with $0\le s \le L$. $L$ is the contour length of the polymer and plays the
role of time.
We assume that the molecules of the polymer repel each other with a
self-avoiding potential $v(\br-\br')$. The potential
$v$ must be strong enough to avoid unwanted intersections of the trajectory
$\br(s)$ with itself.
In the following, the case of two closed or infinitely long polymers
$P_1$ and $P_2$  of contour length $L_1$ and $L_2$ respectively will
be investigated. We suppose that they describe
closed loops or infinitely long Wilson lines $C_1$ and $C_2$
in $3-D$ space.

In order to take into account the entanglement of $P_1$ around $P_2$,
we introduce the Gauss invariant $\chi(C_1,C_2)$:
\begin{equation}
\chi(C_1,C_2)\equiv \frac{1}{4\pi}\oint_{C_1}\oint_{C_2}d\mbd{r}_1\times
d\mbd{r}_2\cdot \frac{{\mbd{r}_1}-\mbd{r}_2}{|\mbd{r}_1 - \mbd{r}_2|^3}
\label{gaussinv}
\end{equation}
Physically, $\chi(C_1,C_2)$ can be interpreted as the potential energy
of a particle  $\br_1$ moving inside a magnetic field $\mbd{B}$
generated by the particle $\br_2$:
\begin{equation}
\chi(C_1,C_2)=\oint_{C_1}d\br_1(s_1)\cdot \mbd{B}(\br_1(s_1))
\label{physgaussinv}
\end{equation}
where

\begin{equation}
\mbd{B}(\br_1)=\frac{1}{4\pi}\oint_{C_2}
d\mbd{r}_2(s_2)\times \frac{{\mbd{r}_1}-\mbd{r}_2(s_2)}{|\mbd{r}_1 -
\mbd{r}_2(s_2)|^3}
\label{magfie}
\end{equation}
For the moment, we confine ourself to the situation in which one single
test polymer, for
instance $P_1$, is entangling with a polymer $P_2$ of static
configuration $\br_2(s_2)$, with $0\le s_2\le L_2$..
The configurational probability $G_m(\br_1,L_1;\br_{0,1},0)$
to find $P_1$ with one end at a point
$\br_1$ starting the other end at $\br_{0,1}$ after intersecting a number
$m=0,1,2,\ldots$ of times an arbitrary surface having as border
$P_2$, can be expressed in terms of
path integrals as the Green function of a particle subjected to a
self-avoiding random walk \cite{tanaka}:
\[
G_m(\br_1,L_1;\br_{0,1},0) =
\int_{\br_{0,1}=\br_1(0)}^{\br_1=\br_1(L_1)}{\cal
D}\br_1(s_1)\delta(\chi(C_1,C_2)-m)\times
\]
\begin{equation}
\mbox{exp}\left \{ -\int_0^{L_1}ds_1{\cal L}_0-\frac{1}{2a^2}
\int_0^{L_1}ds_1\int_0^{L_1}ds_1^\prime
v(\br(s_1)-\br(s_1^\prime))\right\}
\label{singlepoldyn}
\end{equation}
where
\begin{equation}
{\cal L}_0 = \frac 3{2a}\dot{\br}_1^2\label{freelag}
\end{equation}
The case of closed loops or infinitely long polymers is recovered by
taking suitable boundary conditions in the path integral
(\ref{singlepoldyn}). For instance, we require that
$\br_1=\br_{0,1}$ for a closed loop.
To simplify the computations,
it is convenient to work with the chemical potential $\lambda$
conjugated to the topological charge $m$.
Thus we take the Fourier transform of
$G_m(\br_1,L_1;\br_{0,1},0)$ with respect to $m$:
\begin{equation}
G_m(\br_1,L_1;\br_{0,1},0)=
\int \frac{d\lambda}{2\pi}e^{-i\lambda m}
G_\lambda(\br_1,L_1;\br_{0,1},0)\label{ftongm}
\end{equation}
Comparing with eq. (\ref{singlepoldyn}), the Green function
$G_\lambda(\br_1,L_1;\br_{0,1},0)$ is given by:
\[
G_\lambda(\br_1,L_1;\br_{0,1},0)=\int_{\br_{0,1}}^{\br_1}{\cal D}\br_1(s_1)
\mbox{exp}\left \{ -\int_0^{L_1}ds_1{\cal L}_0\right\}\times
\]
\begin{equation}
\mbox{exp}\left \{-\frac{1}{2a^2}
\int_0^{L_1}ds_1\int_0^{L_1}ds_1^\prime
v(\br(s_1)-\br(s_1^\prime))+ i\lambda \chi(C_1,C_2)\right\}\label{aa}
\end{equation}
Let us notice that the the above path integral does not describe a
Markoffian random walk
due to the presence of the non-local self-avoiding term.
To reduce it to a Markoffian random walk, we introduce
gaussian scalar fields $\phi(\br)$ with propagator:
\begin{equation}
\langle \phi(\br)\phi(\br')\rangle =\frac 1{a^2}v(\br -\br')
\label{gpropone}
\end{equation}
Thus we have from eq. (\ref{aa})
\begin{equation}
G_\lambda(\br_1,L_1;\br_{0,1},0)=\langle
G_\lambda(\br_1,L_1;\br_{0,1},0|\phi,\mbd{B})\rangle_\phi\label{aaa}
\end{equation}
In (\ref{aaa}) the symbol
$\langle\enskip\rangle_\phi$ denotes average over the
auxiliary fields $\phi$ and
\begin{equation}
G_\lambda(\br_1,L_1;\br_{0,1},0|\phi,\mbd{B})=
\int_{\br_{0,1}}^{\br_1}{\cal D}\br_1(s_1)
\mbox{exp}\left \{ -\int_0^{L_1}ds_1\left ({\cal L}_\phi+ i 
\dot{\br_1}(s_1)\cdot\lambda\mbd{B}(\br_1(s_1))\right)\right \}\label{bbb}
\end{equation}
where we have put
\begin{equation}
{\cal L}_\phi = {\cal L}_0 + i\phi(r_1(s_1))
\label{ccc}
\end{equation}
We remark that the Green function
$G_\lambda(\br_1,L_1;\br_{0,1},0|\phi,\mbd{B})$ is formally that of a particle
diffusing under the magnetic field $\mbd{B}$ defined in eq. (\ref{magfie})
and an imaginary electric potential $i\phi$.
Thus it can be shown to satisfy the  Schr\"odinger-like equation:
\begin{equation}
\left\{\frac{\partial}{\partial L_1} -\frac a 6({\mbld \nabla}+i\lambda\mbd{B}
)^2+i\phi\right\}G_\lambda(\br_1,L_1;\br_{0,1},0|\phi,\mbd{B})=
\delta(L_1)\delta(\br_1-\br_{0,1})
\label{ddd}
\end{equation}
The Laplace transformed of the above equation with respect to the
contour length $L_1$ is
\begin{equation}
\left\{z_1 -\frac a 6({\mbld \nabla}+i\lambda\mbd{B}
)^2+i\phi\right\}G_\lambda(\br_1, \br_{0,1};z_1|\phi,\mbd{B})=
\delta(\br_1-\br_{0,1})
\label{lapltransf}
\end{equation}
where
\begin{equation}
G_\lambda(\br_1, \br_{0,1};z_1|\phi,\mbd{B}) =
\int_0^\infty dL_1e^{-z_1L_1}G_\lambda(\br_1,L_1;\br_{0,1},0|\phi,\mbd{B})
\label{defla}
\end{equation}
The variable $z_1$ plays the role of the chemical potential conjugate
to the contour length $L_1$. From now on, we set
$\mbd{D}=\mbld{\nabla}
+i\lambda \mbd{B}$.
Starting from eq. (\ref{lapltransf}) and integrating over the
auxiliary fields $\phi$ by means of
the replica trick, one can express the Laplace
transformed $G_\lambda(\br_1,\br_{0,1};z_1)$ of the Green function
(\ref{aa}) in term of second quantized fields.
Skipping the details of the derivation, that can be found
in ref. \cite{tanaka}, we just state the result that one obtains in the
case of a self-avoiding potential of the kind $v(\br) =a^2v_0\delta(\br)$:
\begin{equation}
G_\lambda(\br_1,\br_{0,1};z_1)=\lim_{n\to 0}\int\prod_{\omega=1}^n
{\cal D}\psi^{*\omega}
{\cal D}\psi^{\omega}\psi^{*\overline \omega}(\br_1)
\psi^{\overline \omega}(\br_{0,1})
e^{-F[\Psi]}
\label{onepolymerfinal}
\end{equation}
In the above equation the fields $\psi^{*\omega},\psi^\omega$,
$\omega=1,\ldots, n$,
are complex replica fields and
$\Psi = (\psi^1,\ldots,\psi^n)$.
Moreover,
${\overline \omega}$
is an arbitrarily chosen integer in the range $1,...,n$ and
the polymer free energy $F[\Psi]$ is given by:  
\begin{equation}
F[\Psi] = \int d^3r\left\{\frac a 6||{\mbd D}\Psi||^2+z_1||\Psi||^2
+v_0||\Psi||^4\right\}
\label{spfreeenergy}
\end{equation}
where $||{\mbd D}\Psi||^2=\sum_\omega({\mbd D}\psi^\omega)^\dagger{\mbd D}
\psi^\omega$
and $||\Psi||^2=\sum_\omega|\psi^\omega|^2$.
The generalization of the above formulas
to an arbitrary number $M$ of static polymers
$P_2,\ldots,P_M$ is straightforward, but not the inclusion of their
fluctuations.
Already in the case of two polymers, the analogous of the 
Schr\"odinger equation (\ref{lapltransf}) becomes complicated due to the
presence of the nontrivial interactions between the polymers
introduced by the Gauss invariant (\ref{gaussinv}).
This makes the derivation of the Green function
$G_\lambda(\br_1,L_1;\br_{0,1},0|\phi,\mbd{B})$ in terms of second quantized
fields extremely difficult, apart from a few cases
in which the trajectories of the polymers are strongly constrained.
On the other side, without including the fluctuations of all polymers,
there is always the difficulty of determining those
configurations of the static polymers which are physically relevant.
Indeed, we see from eqs. (\ref{onepolymerfinal}--\ref{spfreeenergy})
that the free energy $F[\Psi]$ of the test polymer $P_1$ depends on the
trajectory $\br_2(s)$
through the magnetic potential ${\mbd B}$ contained in
the covariant derivative $\mbd{D}$.
As we will see in the next section, the introduction of auxiliary Chern--Simons
fields will remove all these problems.

\section{Topological Entanglement of Polymers via Chern--Simons Fields}
We study in this Section the fluctuations of two polymers $P_1$ and $P_2$
subjected to topological constraints. In analogy with
the previous Section, we consider the
configurational probability $G_m(\{\br\},\{L\};\{\br_0\},0)$
of finding the polymer $P_1$ with ends in $\br_1$ and $\br_{0,1}$
and the polymer $P_2$ with ends in $\br_2$ and $\br_{0,2}$.
Moreover, we require that $P_1$ intersects $P_2$ $m$ times.
Here we have put $\{\br\}=\br_1,\br_2$, $\{L\}=L_1,L_2$ etc.
From now on, the indices $\tau,\tau',.... = 1,2$ will be used
to distinguish the two different polymers.
The self-avoiding potential of the previous Section must be extended
in the
present case in order to take into account the reciprocal repulsions
among the molecules of the two different polymers.
Thus we choose a self-avoiding potential of the kind:
\begin{equation}
v_{\tau\tau'}(\br_\tau(s_\tau)-\br_{\tau'}(s'_{\tau'})) =
v^0_{\tau\tau'}v(\br_\tau(s_\tau)-\br_{\tau'}(s'_{\tau'}))
\label{tpselfavpot}
\end{equation}
where $v^0_{\tau\tau'}$ is a symmetric $2\times 2$ matrix and
$v(\br)$ is a strongly repulsive potential.
As seen in the previous Section, the presence of self-avoiding potentials
of this kind leads to random walks which are not Markoffian. To solve this
problem, we introduce here auxiliary scalar fields with gaussian
action and propagator:
\begin{equation}
\langle\phi_\tau(\bx)\phi_{\tau'}(\by)\rangle =\frac 1 {a^2}
v^0_{\tau\tau'}v(\bx -\by)\label{modpot}
\end{equation}
In future we will make use of the following formula:
\[
\int \prod_{\tau=1}^2{\cal D}\phi_\tau
\mbox{\rm exp}
{\left\{-\frac{a^2}2\int d^3\bx d^3\by\phi_\tau(\bx)M^{\tau\tau'}
(\bx,\by)\phi_{\tau'}(\by)-i\sum_{\tau=1}^2\int d^3\bx J_\tau(\bx)
\phi_\tau(\bx)\right\}}=
\]
\begin{equation}
\mbox{\rm exp}\left\{-\frac 1{2a^2}\int_0^{L_\tau}\int_0^{L_{\tau'}}
ds_\tau ds'_{\tau'} v_{\tau\tau'}(\br_\tau(s_\tau)-\br_{\tau'}(s_{\tau'}))
\right\}
\label{eee}
\end{equation}
where $M^{\tau\tau'}(\bx,\by)$ is the inverse of the matrix
$v_{\tau\tau'}(\bx-\by)$ and
\begin{equation}
J_\tau(\bx)=\int_0^{L_\tau}ds_\tau\delta^{(3)}(\bx-\br_\tau(s_\tau))
\label{fff}
\end{equation}
Let us now
rewrite the topological contribution in the
path--integral (\ref{aa}) in a more suitable way by means of
auxiliary C--S
fields.
With the introduction of these fields,
our treatment of the polymer entanglement problem
departs from that of Section 2 and from ref. \cite {tanaka}.

We will consider for our purposes abelian C-S field theories of action:
\begin{equation}
{\cal A}_{CS}(A_\mu,\kappa) =\frac\kappa{8\pi}\int
d^3\bx\epsilon^{\mu\nu\rho}A_\mu \partial_\nu A_\rho\label{csaction}
\end{equation}
with $\mu,\nu,\rho=1,2,3$. $\kappa$ is a real coupling constant and
$\epsilon^{\mu\nu\rho}$ is the completely antisymmetric tensor in
$3-D$. The above action can also be written in another useful form:
\begin{equation}
{\cal A}_{CS}(\bA,\kappa) =\frac\kappa{8\pi}\int
d^3\br \bA \cdot(\mbld{\nabla}\times \bA)\label{pcsaction}
\end{equation}
where $\br=(x^1,x^2,x^3)$.
To quantize the C--S theory we choose the Feynman gauge with propagator:
\begin{equation}
G_{\mu\nu}(\bx,\by)=\frac i\kappa \epsilon_{\mu\nu\rho}
\frac{(x-y)^\rho}{|\bx -\by|^3}
\label{csprop}
\end{equation}
The observables of the theory are gauge invariant operators built out
of the basic fields $A_\mu$. A complete set is given by the holonomies
around closed curves 
\begin{equation}
{\cal W}(C,\gamma)\equiv \mbox{\rm exp} \left\{-i\gamma\oint_C A_\mu dx^\mu
\right\}
\label{holonomy}
\end{equation}
The vacuum expectation value of two of these observables ${\cal
W}(C_1,\gamma_1)$ and ${\cal W}(C_2,\gamma_2)$ is:
\begin{equation}
\langle {\cal
W}(C_1,\gamma_1){\cal W}(C_2,\gamma_2)\rangle_A =
\mbox{\rm exp}\left\{ -i\left(\frac {2\pi}\kappa\right)
\left[\gamma_1^2\chi(C_1,C_1)+\gamma_2^2\chi(C_2,C_2)+2
\gamma_1\gamma_2\chi(C_1,C_2)\right]\right\}
\label{wleval}
\end{equation}
where $\chi(C_\tau,C_\tau)$, $\tau=1,2$ is the so-called
self-linking number of the loop $C_\tau$.

To reproduce the term of eq. (\ref{aa}) containing the Gauss invariant $\chi$,
we need two Chern--Simons
fields $a_\mu$ and $b_\mu$
with actions
${\cal A}_{CS}(\ba,\kappa)$ and ${\cal A}_{CS}(\bb,-\kappa)$
respectively.
Using eq. (\ref{wleval}) and setting for instance
\begin{equation}
\gamma_1=\frac\kappa{4\pi}\qquad \qquad \gamma_2 = \frac \lambda 4
\label{positiona}
\end{equation}
one sees in fact that
\begin{equation}
\langle{\cal W}(C_1,\gamma_1){\cal W}(C_2,\gamma_2)\rangle_a\enskip
\langle{\cal W}(C_1,-\gamma_1){\cal
W}(C_2,\gamma_2)\rangle_b=
\mbox{\rm exp}\left\{-i\lambda\chi(C_1,C_2))\right\}
\label{selflinkelim}
\end{equation}
The right hand side of the above equation is exactly
the contribution due to the topological
entanglements of the polymers appearing in eq. (\ref{aa}).
We are now ready to write the expression of the Green function
$G_\lambda(\{\br\},\{L\};\{\br_0\},0)$ for two entangling polymers.
First of all, let us put:
\begin{equation}
G_\lambda(\{\br\},\{L\};\{\br_0\},0) =\langle
G_\lambda(\{\br\},\{L\};\{\br_0\},0|\{\phi\},\{\bA\})\rangle_{\{\phi\},\ba,\bb}
\label{ggg}
\end{equation}
where $\langle\enskip\rangle_{\{\phi\},\ba,\bb}$
denotes the average with respect
to the fields $\phi_\tau,\ba,\bb$ and
\[
G_\lambda(\{\br\},\{L\};\{\br_0\},0|\{\phi\},\{\bA\})=
\prod_{\tau=1}^2
\int_{\br_{0,\tau}}^{\br_\tau}
{\cal D}
\br_\tau(s_\tau)\mbox{\rm exp}
\left\{-\int_0^{L_\tau}{\cal L}_{\phi_\tau}\right\}\times
\]
\begin{equation}
\mbox{\rm exp}\left\{-i\gamma_\tau\int_0^{L_\tau}\bA^\tau(\br_\tau(s_\tau))
\cdot d\br_\tau(s_\tau)\right\}
\label{hhh}
\end{equation}
The parameters $\gamma_\tau$ appearing in the above equation
are defined as in eq.
(\ref{positiona}) and the fields $\bA^\tau$ are given by the relation:
\begin{equation}
\bA^\tau=\ba+(-1)^\tau\bb\qquad\qquad\tau=1,2
\label{ataudef}
\end{equation}
Exploiting formulas (\ref{eee}) and (\ref{wleval}) in order to
perform the two
independent integrations over the fields $\phi_\tau,\ba$ and $\bb$
in eq. (\ref{ggg}),
one finds that:
\[
G_\lambda(\{\br\},\{L\};\{\br_0\},0) =
\]
\[
\int_{\br_{0,1}}^{\br_1}{\cal D}\br_1(s_1)\int_{\br_{0,2}}^{\br_2}{\cal D}
\br_2(s_2)\mbox{\rm exp}\left\{-\int_0^{L_1}ds_1{\cal L}_0(\dot{\br}_1(s_1))
-\int_0^{L_2}ds_2{\cal L}_0(\dot{\br}_2(s_2))\right\}\times
\]
\begin{equation}
\mbox{\rm exp}
\left\{
-\frac 1 {2a^2}
\sum_{\tau,\tau'=1}^2
\int_0^{L_\tau}
ds_\tau\int_0^{L_{\tau'}}
ds_{\tau'}v_{\tau\tau'}
(\br_\tau(s_\tau)-\br_{\tau'}(s_{\tau'}))-i\lambda\chi(C_1,C_2)\right\}
\label{glamsn}
\end{equation}
By inverse Fourier transformation in $\lambda$ as in
eq. (\ref{ftongm}),
we obtain from eq. (\ref{glamsn}):
\[
G_m(\{\br\},\{L\};\{\br_0\},0) =
\]
\[
\int_{\br_{0,1}}^{\br_1}{\cal D}\br_1(s_1)\int_{\br_{0,2}}^{\br_2}{\cal D}
\br_2(s_2)\mbox{\rm exp}\left\{-\int_0^{L_1}ds_1{\cal L}_0(\dot{\br}_1(s_1))
-\int_0^{L_2}ds_2{\cal L}_0(\dot{\br}_2(s_2))\right\}\times
\]
\begin{equation}
\mbox{\rm exp}
\left\{
-\frac 1 {2a^2}
\sum_{\tau,\tau'=1}^2
\int_0^{L_\tau}
ds_\tau\int_0^{L_{\tau'}}
ds_{\tau'}v_{\tau\tau'}
(\br_\tau(s_\tau)-\br_{\tau'}(s_{\tau'}))\right\}\delta(\chi(C_1,C_2)-m)
\label{gglamsn}
\end{equation}
This is the desired
generalization of eq. (\ref{singlepoldyn}) to the case of two
fluctuating polymers. In fact, if
we ignore the fluctuations of $P_2$ and the reciprocal
repulsion among the molecules of $P_1$ and $P_2$, which was not taken into
account in Section 2, eq. (\ref{glamsn}) exactly
coincides with eq. (\ref{singlepoldyn}).
The advantage of having coupled the polymers with the
Chern--Simons fields is that now each polymer $P_1$ and
$P_2$ undergoes an independent random walk.
Their mutual interaction, that in eqs. (\ref{singlepoldyn}) and
(\ref{gglamsn}) occurred through the Gauss invariant $\chi(C_1,C_2)$, is
now mediated by the Chern--Simons fields, as it is possible to see from
eqs. (\ref{ggg}--\ref{hhh}).
Let us now express the Green function
$G_\lambda(\{\br\},\{L\},\{\br_0\},0)$ in terms of second quantized fields.
To this purpose, we split  the Green function
$G_\lambda(\{\br\},\{L\},\{\br_0\},0|\{\phi\},\{\bA\})$ of
eq. (\ref{hhh}) as follows:
\begin{equation}
G_\lambda(\{\br\},\{L\};\{\br_0\},0|\{\phi\},\{\bA\})=
G_\lambda(\br_1,L_1;\br_{0,1},0|\phi_1,\bA^1)
G_\lambda(\br_2,L_2;\br_{0,2},0|\phi_2,\bA^2)
\label{iii}
\end{equation}
where, for $\tau=1,2$:

\begin{equation}
G_\lambda(\br_\tau,L_\tau;\br_{0_\tau},0|\phi_\tau,\bA^\tau)=
\int_{\br_{0,\tau}}^{\br_\tau}{\cal D}\br_\tau(s_\tau)
\mbox{\rm exp}
\left\{-\int_0^{L_\tau}\left[{\cal L}_{\phi_\tau}-i\gamma_\tau
\bA^\tau\cdot d\br_\tau(s_\tau)\right]\right\}
\label{jjj}
\end{equation}

Each Green function
$G_\lambda(\br_\tau,L_\tau;\br_{0,\tau},0|\phi_\tau,\bA^\tau)$,
$\tau=1,2$, is the Green function of a particle diffusing under the
vector potential $\bA^\tau$ and the imaginary electromagnetic field
$\phi_\tau$. As a consequence, it can be written as the solution
of a Schr\"odinger-like equation like (\ref{ddd}).
In analogy with
the previous Section, it is convenient to introduce the chemical
potentials conjugate to $L_\tau$ and to consider the Laplace
transformed Green function:
\begin{equation}
G_\lambda(\{\br\},\{\br_0\};\{z\}|\{\phi\},\{\bA\})=
\int_0^\infty\int_0^\infty dz_1dz_2e^{-(z_1L_1+z_2L_2)}
G_\lambda(\{\br\},\{L\};\{\br_0\},0|\phi_\tau,\bA^\tau)
\label{tvlaplt}
\end{equation}
From eqs. (\ref{iii})--(\ref{jjj}) we have:
\begin{equation}
G_\lambda(\{\br\},\{\br_0\};\{z\}|\{\phi\},\{\bA\})=
G_\lambda(\br_1,\br_{0,1};z_1|\phi_1,\bA^1)
G_\lambda(\br_2,\br_{0,2};z_2|\phi_2,\bA^2)
\label{glzexp}
\end{equation}
where the functions
$G_\lambda(\br_\tau,\br_{0,\tau};z_\tau|\phi_\tau,\bA^\tau)$
have already been defined in eq. (\ref{defla}).
For each value of $\tau=1,2$, they explicitly satisfy eq.
(\ref{lapltransf}), which we rewrite here for convenience:
\begin{equation}
\left\{z_\tau -\frac a 6\mbd{D}_\tau^2+i\phi_\tau\right\}
G_\lambda(\br_\tau, \br_{0,\tau};z_\tau|\phi_\tau,\mbd{A}^\tau)=
\delta(\br_\tau-\br_{0,\tau})
\label{tplapltransf}
\end{equation}
The covariant derivatives $\mbd{D}_\tau$ are defined as follows:
$\mbd{D}_\tau=\mbld{\nabla}+i\gamma_\tau \bA^\tau$.
The solution of eq. (\ref{tplapltransf}) in terms of complex scalar fields
$\psi^*_\tau,\psi_\tau$ is:
\begin{equation}
G_\lambda(\br_\tau, \br_{0,\tau};z_\tau|\phi_\tau,\mbd{A}^\tau)=
\frac 1{Z_\tau}\int{\cal D}\psi^*_\tau{\cal D}\psi_\tau
\psi^*_\tau(\br_\tau)\psi_\tau(\br_{0,\tau})e^{-F[\psi_\tau]}
\label{kkk}
\end{equation}
where, setting $|\mbd{D}_\tau\psi_\tau|^2=
(\mbd{D}_\tau\psi_\tau)^\dagger\cdot\mbd{D}_\tau\psi_\tau$, the free
energy $F[\psi_\tau]$ is given by:
\begin{equation}
F[\psi_\tau]=\int d^3\br\left[
\frac a 6|{\mbd D}_\tau\psi_\tau|^2 + (z_\tau+i\phi_\tau)|\psi_\tau|^2
\right]
\label{freeenergy}
\end{equation}
Finally, the partition function $Z_\tau$ is:
\begin{equation}
Z_\tau=\int{\cal D}\psi^*_\tau{\cal D}\psi_\tau
e^{-F[\psi_\tau]}
\label{mmm}
\end{equation}
As we see from eq. (\ref{freeenergy}),
$F[\psi]$ is nothing but the Gintzburg-Landau free energy of a superconductor
in a fluctuating magnetic field.
We are now ready to perform the average over the auxiliary fields
$\phi_\tau$ in the Green function (\ref{glzexp}).
This integration is however higly non trivial. As a matter of fact, using
eq. (\ref{kkk}) to express the original Green function
(\ref{glzexp})
in terms of second quantized fields, one immediately
realizes that the integrand is
not gaussian due to the presence of the factors $Z_\tau^{-1}$.
To solve this problem, we exploit the replica trick.
Thus we introduce $2n$ replica
fields $\psi_\tau^{* \omega},\psi_\tau^\omega$, with $\tau=1,2$ and
$\omega=1,\ldots,n$.
In terms of these fields, the Green function (\ref{glzexp})
can be written as follows:
\begin{equation}
G_\lambda(\{\br\},\{\br_0\};\{z\}|\{\phi\},\{\bA\})=
\lim_{n\to 0}\prod_{\tau=1}^2\left[\int\prod_{\omega=1}^n
{\cal D} \psi_\tau^{*\omega}
{\cal D} \psi_\tau^{\omega}\psi_\tau^{*\overline{\omega}}(\br_\tau)
\psi_\tau^{\overline{\omega}}(\br_{0,\tau})e^{-F[\psi_\tau^\omega]}\right]
\label{replicatrick}
\end{equation}
where $\overline{\omega}$ is an arbitrary integer chosen in the range
$1\le \overline{\omega}\le n$.
According to the replica trick, we will also assume that the limit for
$n$ going to zero commutes with the integrations in the fields
$\bA^\tau$ and $\phi_\tau$. In this way, the integral over the
auxiliary
fields $\phi_\tau$ becomes gaussian and can be easily performed.
Supposing in analogy with the previous Section that the self-avoiding
potential is of the kind:
\begin{equation}
v_{\tau\tau'}(\br)=v^0_{\tau\tau'}\delta(\br)
\label{sapot}
\end{equation}
we have after some calculations that:
\[
\langle
G_\lambda(\{\br\},\{\br_0\};\{z\}|\{\phi\},\{\bA\})\rangle_{\{\phi\},
\ba,\bb}=
\]
\begin{equation}
\lim_{n\to 0}\int{\cal D}\ba{\cal D}\bb
\prod_{\tau=1}^2\left[\prod_{\omega=1}^n
{\cal D} \psi_\tau^{*\omega}
{\cal D} \psi_\tau^{\omega}
\psi_\tau^{*\overline{\omega}}(\br_\tau)
\psi_\tau^{\overline{\omega}}(\br_{0,\tau})\right]\mbox{\rm exp}
\left\{-{\cal A}(\ba,\bb,\{\Psi\})\right\}
\label{finalf}
\end{equation}
where, using the same notation of
eqs. (\ref{onepolymerfinal}--\ref{spfreeenergy}), the action ${\cal
A}(\ba,\bb,\{\Psi\})$ is:
\[
{\cal A}(\ba,\bb,\{\Psi\})=i{\cal A}_{CS}(\ba,\kappa)+
i{\cal A}_{CS}(\bb,-\kappa)+
\]
\begin{equation}
\sum_{\tau=1}^2
\left[\frac
{a}6||\mbd{D}_\tau\Psi_\tau||^2+z_\tau||\Psi_\tau||^2\right]
+\sum_{\tau,\tau'=1}^2||\Psi_\tau||^2v^0_{\tau\tau'}||\Psi_{\tau'}||^2
\label{actionf}
\end{equation}
Eq. (\ref{finalf}) is the generalization of
eq. (\ref{onepolymerfinal}), describing in terms
of fields the configurational probability for
two entangling polymers $P_1$ and $P_2$ to have their ends
in $\br_\tau$ and $\br_{0,\tau}$ for $\tau=1,2$ respectively.
This probability is given in the space of the chemical potentials
$\lambda$ and $z_\tau$ conjugated to the topological number $m$ and
the contour lengths $L_\tau$.
It is also possible to find an
expression of the above configurational probability in the space of the
topological number $m$
by taking the inverse Fourier transformation of the
Green function (\ref{finalf}):
\begin{equation}
G_m(\{\br\},\{\br_0\},\{z\})=
\int \frac{d\lambda}{2\pi}e^{-i\lambda m}
\langle
G_\lambda(\{\br\},\{\br_0\},\{z\}|\{\phi\},\{\bA\})\rangle_{\{\phi\},
\ba,\bb}
\label{ftinv}
\end{equation}
To this purpose, we split the action (\ref{actionf})
into three parts:
\begin{equation}
{\cal A}(\ba,\bb,\{\Psi\})=
{\cal A}_0(\{\bA\},\{\Psi\})+\lambda
\int d^3\br\enskip\bi_2(\br)\cdot \bA^2(\br)+
\frac a 6\lambda^2\int \frac{d^3\br}{16} \enskip
\bA^2\cdot \bA^2||\Psi_2
(\br)||^2\label{daction}
\end{equation}
where ${\cal A}_0(\{\bA\},\{\psi^\omega\})$ is the contribution to the action
${\cal A}(\ba,\bb,\{\psi^\omega\})$ which does not contain $\lambda$:
\[
{\cal A}_0(\{\bA\},\{\Psi\})=
i{\cal A}_{CS}(\ba,\kappa)+
i{\cal A}_{CS}(\bb,-\kappa)+
\]
\begin{equation}
\frac{a}6||\mbd{D}_1\Psi_1||^2+
\frac{a}6||\mbld{\nabla}\Psi_2||^2+\sum_{\tau=1}^2
z_\tau||\Psi_\tau||^2
+\sum_{\tau,\tau'=1}^2||\Psi_\tau||^2v^0_{\tau\tau'}||\Psi_{\tau'}||^2
\label{actionzero}
\end{equation}
and
\begin{equation}
\bi_2(\br)=\frac a {12} \frac 1{2i}
\left(\Psi_2^*\nabla\Psi_2 -\Psi_2\nabla\Psi_2^*\right)=
\frac a {12} \frac 1{2i}\sum_{\omega=1}^n
\left(\psi_2^{*\omega}\nabla\psi_2^\omega-
\psi_2\nabla\psi_2^{*\omega}\right)
\label{irrr}
\end{equation}
Performing the Gauss
integral in (\ref{ftinv}) and neglecting irrelevant
constant factors, we have:
\[
G_m(\{\br\},\{\br_0\};\{z\})=
\lim_{n\to 0}\int{\cal D}\ba{\cal D}\bb\prod_{\tau=1}^2\left[
\prod_{\omega=1}^n
{\cal D} \psi_\tau^{*\omega}
{\cal D} \psi_\tau^{\omega}
\psi_\tau^{*\overline{\omega}}(\br_\tau)
\psi_\tau^{\overline{\omega}}(\br_{0,\tau})\right]\times
\]
\begin{equation}
\mbox{\rm exp}\left\{
-{\cal A}_0(\{\bA\},\{\psi^\omega\})\right\}
\mbox{\rm exp}\left\{-\frac 1 4 K^{-1}\left(m-i\sum_{\tau=1}^2
\int d^3\br\enskip\bi_2(\br)\cdot\bA^2(\br)\right)^2\right\}K^{-\frac 1 2}
\label{ffinal}
\end{equation}
where
\begin{equation}
K=\frac a 6\int \frac{d^3\br}{16}\bA^2\cdot\bA^2||\Psi_2||^2
\label{kdef}
\end{equation}
\section{Conclusions}
In this paper a C--S based model of polymers
subjected to topological constraints has been proposed
and applied to the description  of two entangling
polymers $P_1$ and $P_2$.
As we have seen in Section 3, the two polymers, whose action in eq. 
(\ref{gglamsn}) was complicated by the reciprocal interactions introduced
by the Gauss invariant, become completely decoupled after the
introduction of auxiliary C--S
fields (see eq. (\ref{daction})).
Each polymer $P_\tau$, $\tau=1,2$, interacts only with the auxiliary fields
$\bA_\tau,\phi_\tau$ and its action is formally that of a particle
moving in the background of the ``electromagnetic''
field $(\bA^\tau,i\phi_\tau)$.
In this way, the application of the methods explained in Section 2 can also be
extended to the case in which the configurations of both
polymers are non-static.
One advantage of having
included the fluctuations of the second polymer is that the
external magnetic field $\mbd{B}$ of eq.
(\ref{onepolymerfinal}) has been replaced by the quantum C--S fields
of eqs. (\ref{finalf}) and (\ref{ffinal}).
Thus, there is no need to give a physical distribution for the
configurations of the static polymers or to average them using
mean-field type techniques.
On the other side, the complications introduced in our approach by the
fact that both polymers are non-static, are minimal.
The expression of the Green function in eq. (\ref{finalf})
differs from that of eq. (\ref{gglamsn}) only by the presence of two sets
of replica fields instead of one.

The inclusion of the C--S fields in the treatment of
polymers opens the possibility of  taking into account
more sophisticated topological invariants than the Gauss linking number.
For instance, after replacing in our procedure
the fields $\ba$ and $\bb$ with
their nonabelian counterparts, one can obtain
higher order knot invariants from the
radiative corrections of eq. (\ref{selflinkelim}) as shown in ref.
\cite{agua}.
However,
in the nonabelian
case the elimination of the undesired Gauss
self-linking terms occurring in eq. (\ref{selflinkelim}) is valid only at the
first order approximation
with respect to the C--S coupling constant $\kappa$, but not
at higher orders.
A possible solution to this problem is the introduction of
a suitable framing such that 
\begin{equation}
\chi_{framed}(C,C)=0
\label{framedlink}
\end{equation}
On the other side, as already mentioned above, to get rid of
the self-linking contributions with the help of a framing
is not so convenient
as in pure C--S field theories.
In fact, a framing like that in eq.
(\ref{framedlink})
is necessarily depending on the form
of the loop $C$, which in turn is a dynamical variable in the present context.
Thus the choice of a framing would terribly complicate the form of
the Schr\"odinger equation (\ref{tplapltransf}), preventing its solution
in terms of second quantized fields.

Concluding, we have shown here that it is possible to 
couple abelian C--S fields to
polymers avoiding the ambiguous self-linking contributions.
As an example, all the polymer
configurational probabilities derived in \cite{tanaka} for one single
test polymer chain have been generalized to the case of two fluctuating
polymers.
In the future we hope that it will be possible to extend the present approach
also to the case of nonabelian C--S fields and to the situations in which
there is an arbitrary number of polymers. Finally, we remark that our results
could be also applied to other physical systems such as vortex
rings and dislocation lines embedded in a solid, in which topological
contraints among entangled one-dimensional excitations in a
continuum play essential roles.


\begin{thebibliography}{99}

\bibitem{polftmft}
J. Cardy, {\it Phys. Rev. Lett.} {\bf 72} (1994), 1580; M. G. Brereton and
T. A. Vilgis, {\it Jour. Phys. A: Math. Gen.} {\bf 28} (1995), 1149;
S. K. Nechaev and V. G. Rostiashvili, {\it J. Phys. II} {\bf 3} (1993), 91.
\bibitem{polftcsprop} M. Otto and T. A. Vilgis,
{\it J. Phys. A: Math. Gen.} {\bf 29} (1996), 3893;
S. Nechaev, {\it Int. Jour. Mod. Phys.} {\bf B4} (1990), 1809.
\bibitem{gaugemod} M. G. Brereton and S. Shah, {\it J. Phys. A: Math. Gen.} 
{\bf 13} (1980),
2751; {\it ibid.} {\bf 14} (1981), L-51; {\it ibid.} {\bf 15} (1982), 989;
D. J. Elderfield,  {\it J. Phys. A: Math. Gen.} {\bf 15} (1982), 1369.
\bibitem{tanaka} F. Tanaka, {\it Prog. Theor. Phys.} {\bf 68} (1982), 148.
\bibitem{tanakaii} F. Tanaka, {\it Prog. Theor. Phys.} {\bf 68} (1982), 164.
\bibitem{kleinert} H. Kleinert, {\it Path Integrals} (2nd edition),
World Scientific, Singapore, New Jersey, London, Hong Kong 1995.
\bibitem{edwards} S. F. Edwards, {\it Proc. Phys. Soc.} {\bf 91} (1967), 513;
 {\it J. Phys. A: Math. Gen.} {\bf 1} (1968), 15.
\bibitem{degennes} P. G. de Gennes, {\it Phys. Lett.} {\bf A38} (1972), 339;
J. des Cloiseaux, {\it Phys. Rev} {\bf A10} (1974), 1665;
V. J. Emery, {\it Phys. Rev.} {\bf B11} (1975), 239.
\bibitem{kauffmann} L. H. Kauffmann, {\it Knots and Physics}, World Scientific,
Singapore 1993.
\bibitem{witten} E. Witten, {\it Comm. Math. Phys.} {\bf 121} (1989), 351. 
\bibitem{jones} V. F. R. Jones, {\it Bull. Am. Math. Soc.} {\bf 129} (1989),
103.
\bibitem{chernsimons}
R. Jackiw and S. Templeton, {\it Phys. Rev.} {\bf D23} (1981),
2291; S. Deser, R. Jackiw and S. Templeton, {\it Phys. Rev. Lett.}
{\bf 48} (1983), 975;
J. Schonfeld, {\it Nucl. Phys.} {\bf B185} (1981), 157;
C. R. Hagen, {\it Ann. Phys.} (NY) {\bf 157} (1984), 342.
\bibitem{agua}
E. Guadagnini, M. Martellini and M. Mintchev, {\it Nucl. Phys.}
{\bf B336} (1990), 581;
J. M. F. Labastida and A. V. Ramallo, {\it Phys. Lett.}
{\bf 238B} (1989), 214.
\bibitem{volog} A. V. Vologodskii, A. V. Lukashin, M. D. Frank-Kamenetskii
and V. V. Anshelevich, {\it Sov. Phys.} JETP, {\bf 39} (1975), 1059;
{\bf 40} (1975), 1932.
\end{thebibliography}
\end{document}